\documentclass[runningheads]{llncs}
\usepackage[T1]{fontenc}
\usepackage[hidelinks]{hyperref}
\usepackage{tikz}
\usepackage{wrapfig}
\usepackage{caption}
\usepackage{subcaption}
\usepackage{amsmath}
\usepackage{amssymb}
\usepackage{tabularx}
\usepackage{listings}
\usepackage{booktabs}
\usepackage{xcolor}
\usetikzlibrary{automata, positioning, petri}

\usepackage{graphicx}
\begin{document}
\title{Discovering Coordinated Processes From Social Online Networks\thanks{We acknowledge support from the Australian Government through the Australian Research Council’s Discovery Projects funding scheme (project DP210103700).}}
\titlerunning{Discovering Coordinated Processes From Social Online Networks}
%
%
\author{
Anna Kalenkova \inst{1}\orcidID{0000-0002-5088-7602},
Lewis Mitchell\inst{1}\orcidID{0000-0001-8191-1997}, Ethan Johnson\inst{1}\orcidID{0000-0002-2678-3045}}
%
%
\institute{Adelaide Data Science Centre, School of Computer and Mathematical Sciences, \\ The University of Adelaide, North Terrace 5000, Australia \\
\email{\{anna.kalenkova,lewis.mitchell,ethan.johnson\}@adelaide.edu.au}\\
}
\maketitle              
\begin{abstract}
The rapid growth of social media presents a unique opportunity to study coordinated agent behavior in an unfiltered environment. Online processes often exhibit complex structures that reflect the nature of the user behavior, whether it is authentic and genuine, or part of a coordinated effort by malicious agents to spread misinformation and disinformation. 
Detection of AI-generated content can be extremely challenging due to the high quality of large language model-generated text.
Therefore, approaches that use metadata like post timings are required to effectively detect coordinated AI-driven campaigns.
Existing work that models the spread of information online is limited in its ability to represent different control flows that occur within the network in practice. Process mining offers techniques for the discovery of process models with different routing constructs and are yet to be applied to social networks. We propose to leverage process mining methods for the discovery of AI and human agent behavior within social networks.
Applying process mining techniques to real-world Twitter (now~X) event data, we demonstrate how the structural and behavioral properties of discovered process models can reveal coordinated AI and human behaviors online.


\keywords{Social network analysis  \and AI and human online agents \and Process mining \and (Stochastic) Petri nets \and Process discovery.}
\end{abstract}
\section{Introduction}
The rise in popularity of social networks combined with the rapid advancement of artificial intelligence (AI)~\cite{DBLP:journals/corr/abs-2404-03021} has put the spread of misinformation and disinformation at an all-time high~\cite{c2051f521cb94d58abb88f0184b3ad11}. While both misinformation and disinformation pertain to incorrect information, the key difference is that misinformation can be a simple misunderstanding of facts passed as truth while disinformation is deliberately deceptive in nature. Social networks are a perfect breeding ground for this mis/disinformation due to their accessibility and the ease of information spread by coordinated AI and human agents~\cite{Wang2018IsTT}. As such, the discovery of these coordinated processes is a significant issue.

The rise of AI tools like large language models (LLMs) increases the risk of rapid and coordinated mis/disinformation campaigns online, as they make it easier to generate large volumes of realistic text.
A substantial body of research explores AI-generated text detection \cite{kwon2025comprehensive}, and a variety of techniques exist \cite{hu2024bad}.
However, numerous challenges remain when relying solely on text data to detect AI-driven coordinated mis/disinformation campaigns \cite{zeng2024detecting}, and the trend of increasing realism in LLM-generated text means that purely content-based approaches are unlikely to be successful.
Therefore, it is necessary to use metadata from social media posts to detect coordination, with post timings \cite{graham2024coordination} being an attractive approach. 
Process mining thus emerges as a natural framework to detect AI-driven misinformation campaigns.

Several methods have been proposed for discovering coordinated AI behaviors online using metadata from social media posts~\cite{mannocci2024detection}. Recent methods are focused on the discovery and analysis of \emph{(online) social networks} ~\cite{Weber2021ExploringTE},  \emph{coordinated networks}~\cite{10508551}, \emph{synchronized networks}~\cite{Ng2023ACS},  \emph{information cascades}~\cite{CINELLI2022113819}, and \emph{threshold models}~\cite{BOZORGI2017149}. Online social, coordinated and synchronized networks can be modeled as directed or undirected graphs, where nodes represent accounts and edges denote communications. These models are static and do not capture the dynamic nature of social networks. In contrast, information cascades and threshold models represent the concurrent spread of information through a network, but do not account for alternative paths of information propagation.

Another approach to modeling online AI and human behavior is to leverage stochastic processes. Some techniques employ Hawkes processes~\cite{10.1145/3122865.3122874} to model and categorize the user behavior. In these models, an event (such as a post or repost) can trigger new events (reposts) with an intensity that decreases over time. Although they represent the stochastic perspective, Hawkes processes do not capture the structure of the social network. Discrete-time Markov chains~\cite{parmar:2024} and continuous Markov processes~\cite{bolzern:2019} have also been extensively used to discover and analyze user behaviors. While Markov chains and processes can be visualized, these models can be convoluted and challenging to comprehend. Moreover, we will demonstrate that Petri nets are more suitable for modeling concurrent behaviors than Markov processes.

The main motivation for this research is to build models that capture both: (1) the stochastic dynamics of networks, such that the malicious AI behavior can be simulated and analyzed, and (2) the structure of the social network, including parallel and alternative pathways of information spread. Stochastic Petri nets are thus a promising tool for modeling these perspectives together. Although some methods for modeling social networks using Petri nets exist~\cite{KARADOGAN2022100924,7116086}, the discovery and analysis of such models from social network data has not yet been studied.


Seminal work on discovering social networks within the domain of process mining is presented in~\cite{10.1007/978-3-540-25970-1_16}. 
The techniques in~\cite{9477594,10.1007/978-3-031-41620-0_17,DBLP:journals/sosym/NesterovBLP23} focus on discovering agent behavior, while this paper introduces a method for learning process models that capture information flow. Other recent agent mining techniques~\cite{10680657,10680660} discover information flow and identify handovers between agents, but mainly analyze sequential interactions, while we propose an approach for discovering models in which information can be passed to multiple users.

The main contributions of this paper are as follows: (1) an approach for discovering stochastic Petri nets from social network event data; (2) methods for analyzing and categorizing AI and human agent behavior based on the discovered stochastic Petri net models; (3) the application of the proposed techniques to real-world social network data to identify coordinated online behaviors.

 
\section{Motivating Examples}
\label{sec:motiv}
This section justifies the choice of stochastic Petri nets as a modeling tool for representing and analyzing user behavior online. First, we present an example demonstrating how  Petri nets model concurrency in social networks and compare them to Markov processes. Second, we examine threshold models, which can represent concurrency but fail to capture patterns where only a specific group of users interacts with a post.

Consider user $A$ and their followers, users $B$ and $C$ (see the social network presented in~\autoref{fig:social_network}).

\begin{center} 
\vspace{-30pt}
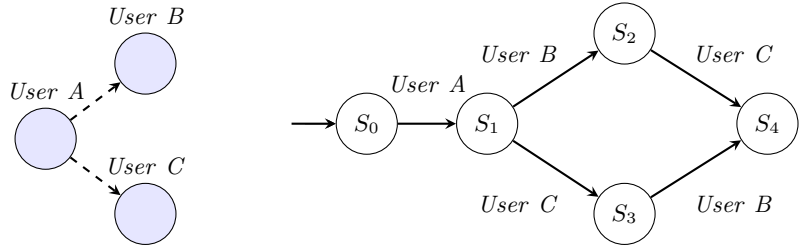
\begin{figure}
    \begin{subfigure}{0.35\textwidth} 
        \centering
        \begin{tikzpicture}[>=stealth, node distance=0.5cm]
            \node[state, fill=blue!10] (A) {\text{ }};
            \node[above=of A, yshift=-0.5cm] {$\mathit{User~A}$};
            \node[state, right=of A, yshift=1cm, fill=blue!10] (B) {};
            \node[above=of B, yshift=-0.5cm] {$\mathit{User~B}$};
            \node[state, right=of A, yshift=-1cm, fill=blue!10] (C) {};
            \node[above=of C, yshift=-0.5cm] {$\mathit{User~C}$};

            \draw[->, thick, dashed] (A) -- (B);
            \draw[->, thick,dashed] (A) -- (C);
        \end{tikzpicture}
        \caption{A social network where users~B and C follow user A.}
        \label{fig:social_network}
    \end{subfigure}
    \hspace{0.2cm}
    \begin{subfigure}{0.45\textwidth} 
        \centering
        \begin{tikzpicture}[>=stealth, node distance=1cm]
            \node[state, initial, initial text=] (Z) {$S_0$};
            \node[state, right of=Z, xshift=0.6cm] (A) {$S_1$};
            \node[state, right=of A, yshift=1.2cm] (B) {$S_2$};
            \node[state, right=of A, yshift=-1.2cm] (C) {$S_3$};
            \node[state, right=of A, xshift=1.9cm] (D) {$S_4$};

            \draw[->, thick] (-1,0) -- (Z);
            \draw[->, thick] (A) -- (B) node[midway, above, xshift=-0.5cm, yshift=0.1cm]{$\mathit{User~B}$};
            \draw[->, thick] (A) -- (C) node[midway, above, xshift=-0.5cm, yshift=-0.7cm]{$\mathit{User~C}$};
            \draw[->, thick] (B) -- (D) node[midway, above, xshift=0.5cm, yshift=0.1cm]{$\mathit{User~C}$};
            \draw[->, thick] (C) -- (D) node[midway, above, xshift=0.5cm, yshift=-0.7cm]{$\mathit{User~B}$};
            \draw[->, thick] (Z) -- (A) node[midway, above, xshift=0cm, yshift=0.3cm]{$\mathit{User~A}$};
        \end{tikzpicture}
        \caption{A Markov chain representation of the social network dynamics.}
        \label{fig:markov_chain}
    \end{subfigure}
    \caption{A social network where users B and C follow user A and the dynamics of the social network.}
    \vspace{-15pt}
    \end{figure}
\end{center}
\vspace{-20pt}

Each time user $A$ posts or reposts information, it is subsequently reposted by $B$ and $C$. The event log generated by this model will contain traces in which activities initiated by users $B$ and $C$ alternate, i.e., $\langle \mathit{User~A}, \mathit{User~B}, \mathit{User~C}\rangle$ and $\langle \mathit{User~A}, \mathit{User~C}, \mathit{User~B}\rangle$. 
\begin{wrapfigure}{l}
{0.45\textwidth} 
\vspace{-20pt}
    \centering
    \begin{tikzpicture}[node distance=0.5
    cm, every place/.style={circle, draw, minimum size=6mm, line width=0.5mm}, every transition/.style={rectangle, draw, minimum height=6mm, minimum width=12mm, fill=red!80!green!40, line width=0.5mm}]

        \node (p1) [place, tokens=1] {};

        \node (tA) [transition, right=of p1] {$\mathit{User~A}$};
        \node (p2) [place, right=of tA, yshift=5mm] {};
        \node (p3) [place, right=of tA, yshift=-5mm] {};
        \node (tB) [transition, right=of p2] {$\mathit{User~B}$};
        \node (tC) [transition, right=of p3] {$\mathit{User~C}$};

        \draw [->, thick] (p1) -- (tA);
        \draw [->, thick] (tA) -- (p2);
        \draw [->, thick] (tA) -- (p3);
        \draw [->, thick] (p2) -- (tB);
        \draw [->, thick] (p3) -- (tC);

    \end{tikzpicture}
    \caption{A Petri net modeling the spread of information from user $A$ to users $B$ and $C$.}
    \label{fig:petri_net}
    \vspace{-25pt}
\end{wrapfigure}
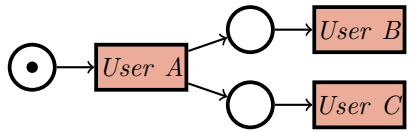
\autoref{fig:markov_chain} illustrates a Markov chain representing the dynamics of the social network, where states denote its configurations and transitions correspond to user activities.
The Petri net, together with its initial marking, that models the spread of information from user $A$ to users $B$ and $C$ is represented in~\autoref{fig:petri_net}. 
This Petri net models concurrency; however, the corresponding Markov chain (\autoref{fig:markov_chain}) assumes that users make reposts sequentially~\footnote{Real-world datasets typically include only the original tweet identifier, and concurrency needs to be discovered.}. As a result, the waiting times for the same dataset will be interpreted as shorter, because, in some cases, users $B$ and $C$ do not wait for $A$ to submit the post but instead wait for each other.

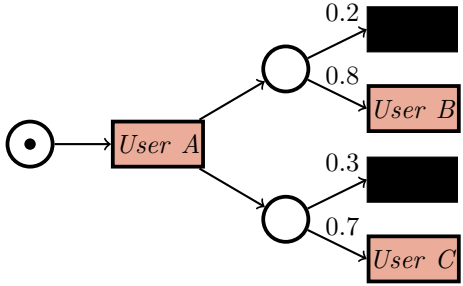
\begin{wrapfigure}{l}{0.5\textwidth} 
\vspace{-0.7cm}
\begin{tikzpicture}[node distance=0.75
    cm, every place/.style={circle, draw, minimum size=6mm, line width=0.5mm}, every transition/.style={rectangle, draw, minimum height=6mm, minimum width=12mm, fill=red!80!green!40, line width=0.5mm}]

        \node (p1) [place, tokens=1] {};

        \node (tA) [transition, right=of p1] {$\mathit{User~A}$};
        \node (p2) [place, right=of tA, yshift=10mm] {};
        \node (p3) [place, right=of tA, yshift=-10mm] {};
        \node (tB) [transition, right=of p2, yshift=-15pt] {$\mathit{User~B}$};
        
        \node [draw, fill=black, minimum width=12mm, minimum height=6mm, right=of p2, yshift=15pt] (tE1){};

        \node [draw, fill=black, minimum width=12mm, minimum height=6mm, right=of p3, yshift=15pt] (tE2){};

        \node (tC) [transition, right=of p3, yshift=-15pt] {$\mathit{User~C}$};

        \draw [->, thick] (p1) -- (tA);
        \draw [->, thick] (tA) -- (p2);
        \draw [->, thick] (tA) -- (p3);
        \draw [->, thick] (p2) -- node[midway, above, xshift=2pt] {0.8} (tB.west);
        \draw [->, thick] (p2) -- node[midway, above, xshift=2pt, yshift=5pt] {0.2}(tE1.west);
        \draw [->, thick] (p3) -- node[midway, above, xshift=2pt] {0.7} (tC.west);
        \draw [->, thick] (p3) -- node[midway, above, xshift=2pt, yshift=5pt] {0.3} (tE2.west);

    \end{tikzpicture}
        



            \caption{A stochastic Petri net for the threshold model. ``Black box'' transitions indicate silent behavior when the post is not reposted by user B or user C.}
\label{fig:petri_net_as_threshold_model}
\vspace{-20pt}
\end{wrapfigure}

Threshold models~\cite{BOZORGI2017149}, in turn, are used to model concurrent processes within social networks. Consider the social network presented in \autoref{fig:social_network}. In addition to the graph structure of the network, threshold models provide probabilities that a post will be reposted by followers. For example, user $B$ reposts posts from user $A$ with probability $0.8$, and user $C$ reposts posts from user $A$ with probability $0.7$. This can be modeled by a stochastic Petri net, as shown in~\autoref{fig:petri_net_as_threshold_model}, where ``black box'' transitions represent silent behavior when the post is not reposted.
Although stochastic Petri nets can model behavior defined by threshold models, they also support choice constructs. This capability is particularly important when different groups of users are interested in specific topics. 

The next section introduces the basic concepts, including free-choice stochastic Petri nets and related structural and behavioral measures, used in this paper. 

\vspace{-5pt}
\section{Theoretical Background}
\label{sec:background}

In this section, we provide the basic definitions used to describe the proposed social network analysis approach.
We first introduce stochastic process models and then define structural and behavioral characteristics of these models.

\subsection{Process Models}

\begin{definition}[Petri net]
\label{def:petri_net}
    Let $\Lambda$ be a set of labels. A Petri net is a 5-tuple $PN = (T, P, F, l, M_0)$, where 
 $T$ is a finite set of transitions,
 $P$ is a finite set of places, such that $P \cap T = \emptyset$,
$F$ is a set of arcs, defined as $F \subseteq (P \times T) \cup (T \times P)$,
 $l: T \to \Lambda \cup \{\tau\}$ is a labeling function that assigns labels to transitions, symbol $\tau\not\in\Lambda$~denotes silent transitions,
 $M_0: P \rightarrow \mathbb{N}$ is the initial marking, which defines the initial number of tokens in each place. For a node $n\in P\cup T$ the set of \emph{input} nodes and set of \emph{output} nodes are defined as $\bullet n=\{x|(x,n)\in F\}$ and $n\bullet =\{x|(n,x)\in F\}$, respectively.
\end{definition}


\autoref{fig:petri_net} illustrates an example of a Petri net. Places are depicted as circles, transitions as rectangles, and arcs as arrows. 
The initial marking of the Petri net is shown as a token in the leftmost place. Tokens move when a transition fires. The formal definition of transition firing is given below.

\begin{definition}[Transition firing]
    \label{def:transition_firing}
   A transition \( t \in T \) is \emph{enabled} in a marking \( M:P\rightarrow \mathbb{N} \) if, for all places \( p \in \bullet t\), it holds that \( M(p) \geq 1 \), i.e., each input place contains at least one token. 
An enabled transition \( t \) can \emph{fire}, consuming one token from each of its input places and producing one token to each of its output places. This results in a new marking \( M' \), defined as follows: 1) \( M'(p) = M(p) - 1 \), if $p\in\bullet t$ and $p\not\in t\bullet$, 2)~\( M'(p) = M(p) + 1 \), if $p\in t\bullet$ and $p\not\in \bullet t$, 3) \( M'(p) = M(p) \), in all other cases.
We denote this firing as: $M\xrightarrow{\Lambda(t)} M'$.

\end{definition}

In this paper, we focus on free-choice Petri nets, which can be discovered using state-of-the-art process discovery techniques and  model choice patterns. 

\begin{definition}[Free-choice Petri net]
\label{def:free-choice}
A Petri net $\mathit{PN} = (T, P, F, l, M_0)$ is called free-choice iff $\forall t_1,t_2\in T$, $t_1\not=t_2$ if $\exists p\in \bullet t_1, p\in \bullet t_2$, then $\not\exists p'\in P$, such that,  $p'\not=p,p'\in \bullet t_1$.
\end{definition}

As stated in~\autoref{def:free-choice}, if two transitions share an input place, they must not have any additional input places. For example, the Petri nets shown in~\autoref{fig:petri_net} and \autoref{fig:petri_net_as_threshold_model} are free-choice because their transitions either do not share input places or share the same single input place.

The behavior of a Petri net can be described by its reachability graph.

\begin{definition}[Reachability graph]
Let $\mathit{PN} = (T, P, F, l, M_0)$ be a Petri net. We say that marking $M_n$ is reachable from marking $M_1$ iff there exists a sequence of transitions $t_1,t_2,\dots, t_{n-1}$, such that, $M_1\xrightarrow{t_1}M_2\xrightarrow{t_2}\dots\xrightarrow{t_{n-1}}M_n$. By $\mathcal{R}(M)$ we will denote the set of all markings reachable from $M$. The reachability graph of $\mathit{PN}$ is a  directed graph $(G,E)$, $G=\mathcal{R}(M_0)$, $E\subseteq G\times(\Lambda\cup\{\tau\})\times G$, where $e=(M_i,\lambda,M_j)\in E$ iff $M_i\xrightarrow{\lambda}M_j$.
\end{definition}

We assume that all reachability graphs are finite, as this is typically the case for Petri nets discovered from event data by most process discovery algorithms.

We extend the concept of a Petri net by introducing arc probabilities and probabilistic delays for transitions, as arc probabilities and time delays will be used to categorize user behavior.
Among all types of stochastic Petri nets, \emph{generally distributed transition stochastic Petri nets} (GDT\textunderscore SPNs)~\cite{10.1007/978-3-319-06257-0_2} and \emph{labeled generalized stochastic Petri nets} (LGSPNs)~\cite{LEEMANS2024102383} support silent transitions. However, these models impose a restriction that silent immediate transitions must fire first. Additionally, in LGSPNs, exponential distributions can be assigned to silent timed transitions. These limitations do not always apply to Petri nets discovered from event logs, where silent transitions are used to model loop and skip patterns, and therefore are not necessarily required to fire first or to represent additional time delays.

These limitations motivate us to propose a new definition of stochastic Petri nets suitable for discovery from event data. Since state-of-the-art process discovery algorithms used in practice, such as Inductive Miner~\cite{10.1007/978-3-319-07734-5_6}, tend to discover free-choice Petri nets, which are easily interpretable and can be represented as flowchart  process diagrams, we adopt a free-choice structure to provide clear semantics for conditions associated with the process branches.
We refer to this class of models as \emph{free-choice stochastic Petri nets}.

\begin{definition}[Free-choice stochastic Petri net]
\label{def:stochastic_pn}
    A 7-tuple $\mathit{FSPN} = (T, P,\allowbreak F, l, M_0, \mathit{Pr},D)$, where
        $(T, P, F, l, M_0)$ is a free-choice Petri net,
        $\mathit{Pr}:F\cap(P\times T)\rightarrow [0,1]$ is the function that maps output arcs of places to probabilities, such that $\forall p\in P$, if $\exists t\in p\bullet$, then $\sum\limits_{t\in p\bullet}{\mathit{Pr}(p,t)}=1$,
        $D:T\backslash\{t\in T:\Lambda(t)=\tau\}\rightarrow \mathcal{X_D}$ is the function that maps not silent transitions to independent random variables from the set $\mathcal{X_D}$ that define random delays\footnote{Time distributions in social networks are  not necessary exponential or gamma but can  also be  heavy-tailed.}.        
\end{definition}


Below, we define the semantics of free-choice stochastic Petri nets in a way analogous to the semantics of timed Petri nets as introduced in~\cite{ZUBEREK1991627}.

\begin{definition}[Transition firing in free-choice stochastic Petri nets]
Consider a free-choice stochastic Petri net $\mathit{FSPN} = (T, P, F, l, M_0, \mathit{Pr}, D)$ in a marking $M$. In this marking, for all places $p$, such that $M(p)>0$, if transitions from $p\bullet=\{t_1,\cdots,t_k\}$ are enabled (\autoref{def:transition_firing}), only one transition $t_j\in p\bullet$ is selected in respect to $Pr(p,t_1),\cdots,Pr(p,t_k)$ probabilities. This transition  $t_j$ consumes input tokens (\autoref{def:transition_firing}) and fires with a delay drawn from distribution $D(t_j)$, if it is not silent, and with 0~delay, if it is silent. Transitions with the same delay can fire in any order with no time delay between the firings.

    
\end{definition}





\subsection{Structure Graph-Based Measures}

Petri nets are graph-based models consisting of nodes and edges. We introduce graph-based measures for analyzing Petri nets, using a directed graph \( G = (V,E) \), where \( V = P \cup T \) is the set of nodes, \( E = F \) is the set of edges.

\begin{definition}[Graph density]
For a graph $G=(V,E)$ with $|E|$ number of edges and $|V|$ number of nodes, the graph density is:
$D(G) = \frac{|E|}{|V|(|V|-1)}.$
\end{definition}

\begin{definition}[Graph diameter]
The diameter of a graph $G=(V,E)$ is the longest shortest path between two nodes:
$
\textrm{d}(G) = \max\limits_{u,v\in V} \{\text{min distance from $u$ to $v$}\}.
$
\end{definition}

\subsection{Behavioral Process Measures}





The first behavioral measure we apply is user waiting times. These times are calculated as the delays from the moment a transition is enabled to the moment it fires and releases tokens to its output places.

The other behavioral measure that we consider is Kolmogorov-Sinai entropy, and it is based on the underlying Markov chain.
Due to the intrinsic behavior with arbitrary delays, free-choice stochastic Petri nets cannot be precisely represented as Markov chains with a finite or even countable number of states. However, stochastic Petri net behavior can be approximated and generalized by a discrete Markov chain built on top of the reachability graph and defined by a stochastic matrix $P=(P_{i,j})$, where each entry $P_{i,j}$ is the probability of transition from marking $i$ to marking~$j$.

Such a Markov chain can be obtained, for example, through the estimation of transition frequencies based on event data, and it will annotate the reachability graph with transition probabilities. 
Once the Markov chain is constructed, we can build its totally connected version by adding transitions that connect end states to start states and are labeled with probability 1~\cite{DBLP:journals/corr/abs-1812-07334}. This will allow us to estimate the variability in model behavior by calculating Kolmogorov-Sinai entropy.

\begin{definition}[Kolmogorov-Sinai entropy]
For a totally connected Markov chain with stochastic transition matrix \( P = (P_{i,j}) \), where \( P_{ij} \) is the transition probability from state \( i \) to state \( j \), let \( \mu_1, \mu_2,\dots, \mu_n \) be stationary distributions of states satisfying:
$\mu_j = \sum_{i\in V} \mu_i P_{ij}, \quad \forall j\in[1,.., n].$ 
The Kolmogorov-Sinai entropy is then given by:  
$
h_{\mathit{KS}}(P) = -\sum_{i \in V} \mu_i \sum_{j \in V} P_{ij} \log P_{ij}.$
\end{definition}

\section{Case Studies}
\label{sec:case_studies}

In this section, 
we analyze Twitter datasets\footnote{\href{https://zenodo.org/records/10650967}{https://zenodo.org/records/10650967}, \href{http://zenodo.org/records/10669936}{http://zenodo.org/records/10669936}.} collected for periods from January 2019 till March 2023 to identify  coordinated efforts of individuals and groups in Honduras, United Arab Emirates~\cite{10508551}, and Brazil~\cite{10.1145/3589334.3645651} to manipulate public discourse\footnote{The code and the data are available at:\\ \href{https://github.com/ethanmjohnson/social_network_processes.git}{https://github.com/ethanmjohnson/social\_network\_processes.git}.}. 
Coordinated and uncoordinated online behavior is represented by different datasets in both  Honduras and United Arab Emirates (UAE) data. For the Brazil 2018 election dataset, we split the event data according to the bot scores~\cite{Wojcik2018} provided. Specifically, posts and reposts made by users with bot scores above 0.9 were added to one dataset (Brazil~1), while those with scores below 0.1 were added to another (Brazil~2). Consequently, we built two datasets from the the Brazil 2018 election data.

The analyzed datasets contain the retweet information of users: the tweet they are sharing (trace identifier), the user that has retweeted the tweet (activity name), and the time at which they retweeted (timestamp). 
This dataset, like most social network datasets, does not contain information about which retweet was reposted. It only includes the original tweet number, meaning there is no information about which user influenced another. In such cases, process discovery algorithms, such as  Inductive Miner~\cite{10.1007/978-3-319-07734-5_6}, can be useful as they can identify concurrency and retweeting patterns. Based on the frequencies of directly-follows relations between the events, these algorithms can distinguish between the cases where two independent users followed the same predecessor and cases where they were  following each other (see the example in~\autoref{sec:motiv}).


We applied Inductive Miner with a noise threshold of 0.2 to all six event logs, corresponding to the three countries and different user behaviors.
To make Inductive Miner applicable\footnote{The experiments were run on an Intel Xeon w5-3435X × 32 with 512 GB of RAM.} to these large datasets, we limited each trace to its first 10 events. This adjustment was acceptable because fewer than 38\% of the traces (except for the Brazil 2 event log) contained 10 or more events. As will be demonstrated later, even the beginning of a social network trace can indicate coordinated behavior and provide its early prediction.
Additionally, we considered the first 300, 400, and 200 traces for the datasets of UAE, Honduras, and Brazil, respectively, because due to the nature of social network data, the event logs and the discovered Petri nets (\autoref{tab:petri_nets}) contained a large number of activities and nodes, even after filtering. As part of future work, we plan to enhance the current discovery technique by developing and applying a divide-and-conquer approach, in which the social network is discovered in parts and the resulting subnets are merged.


To further analyze the characteristics of social processes, we examine the structural and behavioral properties of Petri nets discovered from these datasets. 
Free-choice Petri nets discovered by the Inductive Miner were further enhanced with probabilities and time distributions based on the original event log data. The structural and behavioral characteristics of the discovered free-choice stochastic Petri nets are presented in~\autoref{tab:petri_nets}.
The number of nodes represents the total count of transitions and places. As observed in~\autoref{tab:petri_nets}, these values are quite large compared to the number of nodes in typical business process models discovered from event logs, because they directly reflect the number of activities in the event data, which tends to be higher in social network processes.

\tiny
\begin{table}
    \vspace{-7pt}
    \begin{tabularx}{\linewidth}{>{\arraybackslash}p{3cm}|>{\centering\arraybackslash}p{1.5cm}|>{\centering\arraybackslash}p{1.3cm}|>{\centering\arraybackslash}p{1.3cm}|>{\centering\arraybackslash}p{2.2cm}|>{\centering\arraybackslash}p{2.cm}}
        Event log & \# Nodes & Density & Diameter & Mean waiting time & KS entropy\\ 
         \hline
          \hline
        UAE coord.  & 1,349  & 0.0010  & 59& 42,179& 2.13\\
        \hline
        UAE uncoord.  & 3,385  & 0.0003  & 155& 68,624& 0.46\\ \hline
        Honduras coord.  & 1,128  & 0.0012  & 111& 16,697 & 1.19\\
        \hline
        Honduras uncoord.  & 3,993  & 0.0003  & 165& 61,277& 0.70\\
        \hline
        Brazil 1  & 1,138  & 0.0011 & 97& 67,104& 0.57\\ 
        \hline
        Brazil 2  & 2,628  & 0.0005 & 113& 18,508& 2.19\\
    \end{tabularx}
    \vspace{5pt}
    \caption{Characteristics, including number of nodes, density, diameter, mean user waiting times in seconds, and Kolmogorov-Sinai entropy (KS entropy), of free-choice stochastic Petri nets discovered from the social event data.}
    \label{tab:petri_nets}
    \vspace{-20pt}
\end{table}
\normalsize

\begin{figure}[h!]
    \centering
    \includegraphics[width=0.65\textwidth]{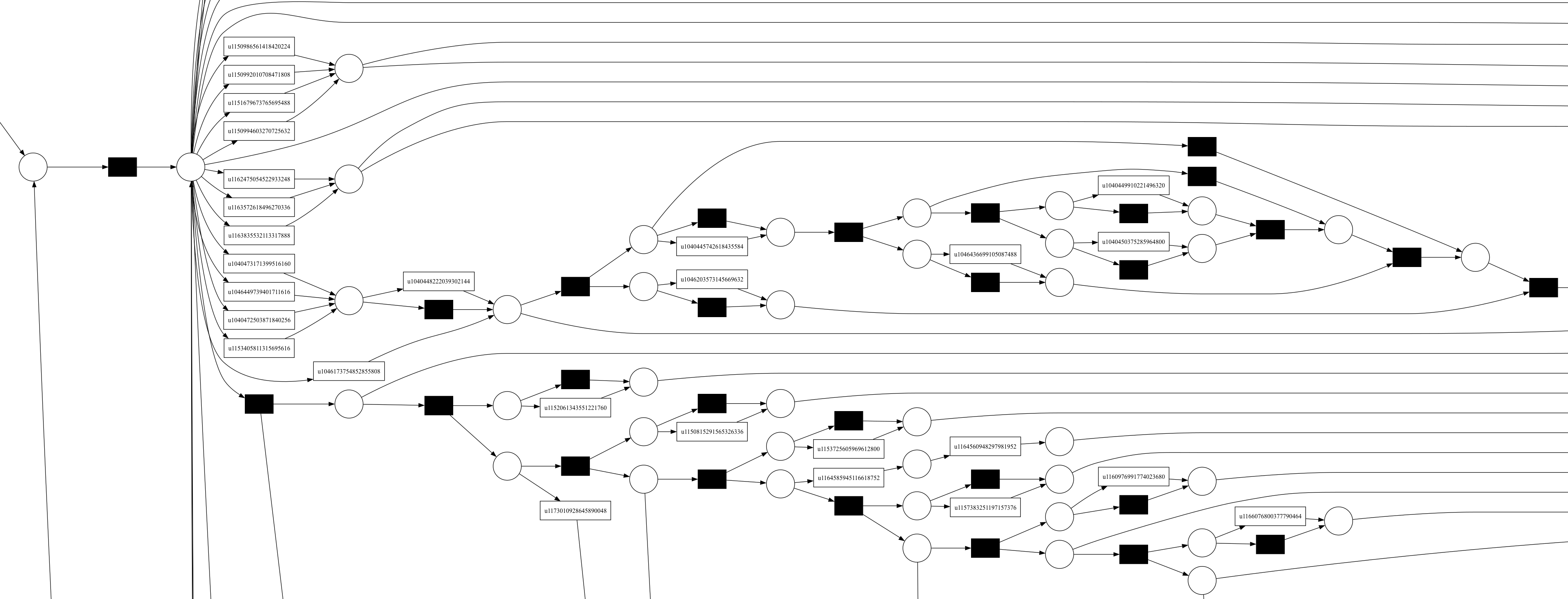} 
    \caption{A fragment of a Petri net discovered from the dataset representing coordinated user behavior in Honduras.}
    \label{fig:coordinated_honduras}
\end{figure}

The density of Petri nets representing uncoordinated behavior is 2 to 4 times lower than that of the corresponding Petri nets modeling coordinated behavior. The density parameter is independent of model size and can be considered an indicator of tighter communication and coordination between the users. A fragment of a Petri net discovered from the event data of coordinated user behavior in Honduras is presented in~\autoref{fig:coordinated_honduras}. This model, for example, contains a large ``flower'' pattern, when several users retweet in any order in a loop, and several parallel, sequential, and choice patterns inside this loop. These patterns represent coordinated groups of users.

\begin{figure}[h!]
    \vspace{-10pt}
    \label{fig:uncoord_honduras}
    \centering
    \includegraphics[width=0.65\textwidth]{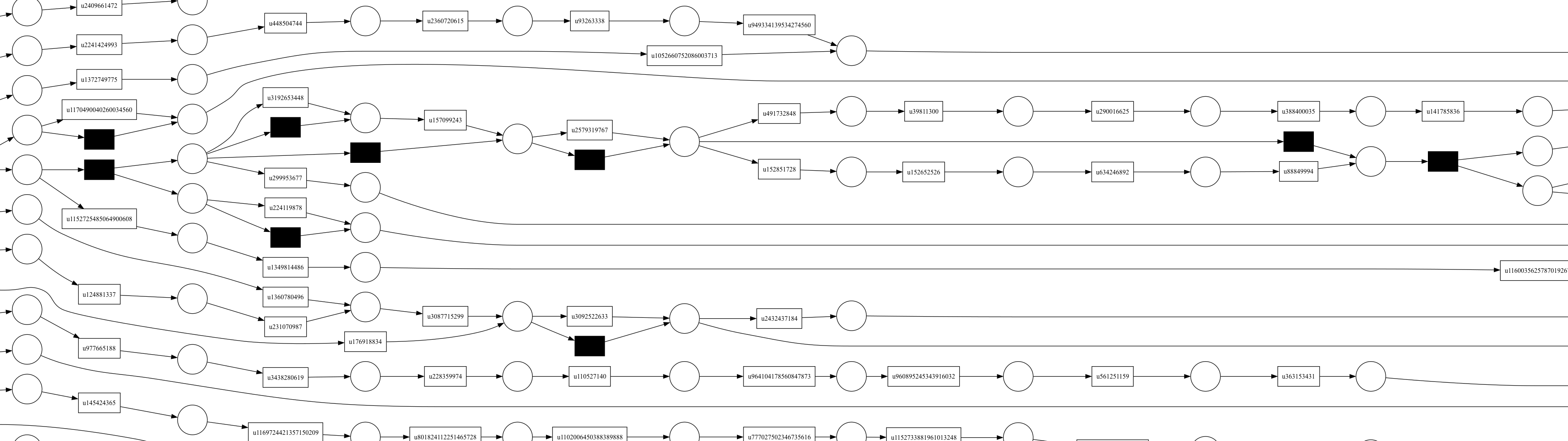} 
    \caption{A fragment of a Petri net discovered from the dateset containing uncoordinated user behavior in Honduras.}
    \label{fig:uncoordinated_honduras}
\end{figure}

\vspace{-20pt}
\autoref{fig:uncoordinated_honduras} shows a typical fragment of the Honduras dataset social network model representing uncoordinated behavior. While this model  contains parallel and choice patterns, users are less interconnected, resulting in a sparser model. This is supported by the density and diameter parameters calculated for the coordinated and uncoordinated models. Overall, social process models representing coordinated behavior tend to be more compact (lower diameter) and denser.

\begin{figure}[h!]
    \begin{subfigure}{0.5\textwidth}
\includegraphics[width=1.0\textwidth]{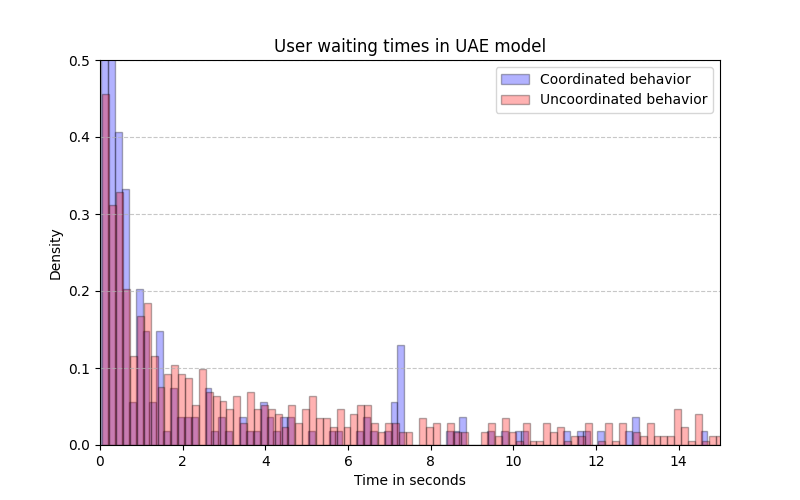}
        \caption{User mean waiting times for UAE coordinated and uncoordinated models.}
        \label{fig:uae_plot}
    \end{subfigure}
    \begin{subfigure}{0.45\textwidth} 
        \centering
    \end{subfigure} 
    \begin{subfigure}{0.5\textwidth}
\includegraphics[width=1.0\textwidth]{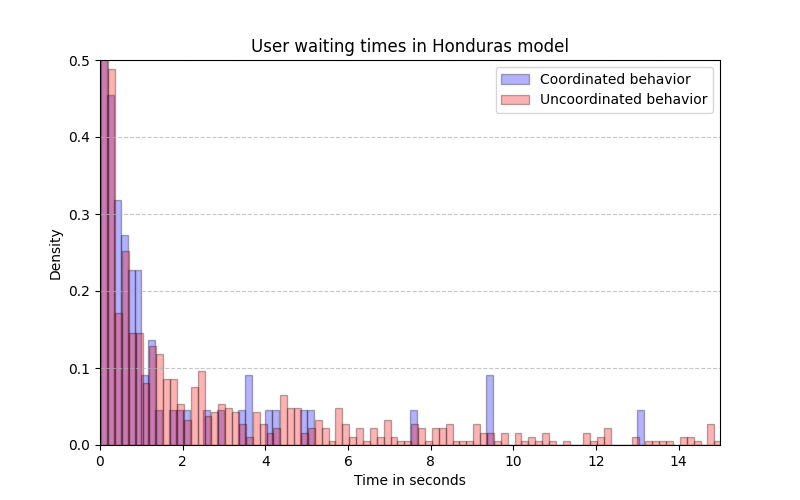}
        \caption{User mean waiting times for Honduras coordinated and uncoordinated models.}
         \label{fig:honduras_plot}
    \end{subfigure}
    \caption{User mean waiting times for process models representing coordinated and uncoordinated behaviors.}
    \vspace{-10pt}
    \end{figure}

In addition to analyzing the structural parameters of the discovered models, we analyzed their behavioral characteristics. First, we assessed user waiting times, which can be calculated as the time gap between the moment a transition is enabled (see~\autoref{def:stochastic_pn}) and the moment it fires, changing the marking. \autoref{tab:petri_nets} contains mean of mean user retweet times (first, the mean waiting time is calculated for each user and then the mean value across all users is estimated). It shows that coordinated users retweet in mean faster than uncoordinated in the UAE and Honduras social process models. \autoref{fig:uae_plot} and    \autoref{fig:honduras_plot} present mean user waiting time distributions for coordinated and uncoordinated behaviors in UAE and Honduras models, respectively.
The Kolmogorov-Smirnov test rejects the hypothesis that coordinated and uncoordinated behaviors are drawn from the same distribution with confidence  $8.82\times 10^{-14}$ and $3.30\times 10^{-10}$, for UAE and Honduras, respectively.

In contrast to the UAE and Honduras social process model results, mean user waiting times for the Brazil social network models exhibit different characteristics. The model discovered from the Brazil~1 dataset with users classified as bots has larger waiting times than the model discovered from the Brazil~2 dataset with users classified as not bots. More detailed analysis of the model discovered from the dataset Brazil~2 shows that this model contains a large ``flower'' loop pattern of more than 200 interconnected users. Additionally, a group of users from Barsil~2 dataset retweet in less than 1 minute or even less than 1 second. According to the study in~\cite{Rauchfleisch2020}, which describes the original Brazil election dataset, Botometer~\cite{Wojcik2018} can assign false negative and false positive botscores
and the results should also be checked manually. Therefore,  the social network models discovered from Brazil~1 and Brazil~2 datasets should be analyzed with careful consideration as models with potentially mixed types of users. 

We calculated Kolmogorov-Sinai entropy of the discovered models based on state frequencies and transition probabilities. As the results suggest, Kolmogorov-Sinai entropy is larger for models built from the UAE and Honduras datasets representing coordinated behavior. This can be attributed to groups of bots retweeting information in  chaotic and unpredictable patterns, leading to increased entropy. However, regular users' behavior is more predictable with lower entropy values. For the Brazil social network models, results align with their time distribution characteristics, the Brazil~1 model exhibits smaller entropy values than the Brazil~2 model.
 
Overall, the structural and behavioral characteristics of discovered social process models consistently characterize bot behavior in UAE and Honduras datasets with established ground truth (i.e., thorough automated and manual analysis and labeling). However, these measures may exhibit inconsistencies, as observed in the case of the Brazil election dataset, where diverse user types may be present. Importantly, the proposed techniques can aid in identifying bot-like behavior within these mixed models by analyzing different sub-parts of the social processes. We consider this a promising direction for future research. 

\vspace{-5pt}
\section{Conclusion}
\label{sec:conclusion}

While various approaches exist for modeling and analyzing user behavior online, process models, including Petri nets, which capture advanced routing patterns such as choice and concurrency, have never been discovered from social network data before. In addition to representing complex routing patterns, these models have formal semantics and can be visualized, making them suitable for both qualitative and quantitative analysis. 


This paper presents a new approach to analyze online user behavior based on process mining techniques that can discover models with advanced routing constructs from event data and distinguish between concurrent and sequential behavior. 
Our approach is effective in practice in categorizing user behavior online and can be applied to identify, visualize and simulate coordinated behavior. In future work, we aim to explore other process discovery algorithms and develop a modular divide-and-conquer technique for discovering Petri nets from large-scale social event data. We also plan to apply the approach to identify bot-like behavior in models with mixed user types.

%

\bibliography{ref.bib}

\end{document}